\documentclass[aps,prb,amssymb,showpacs,superscriptaddress,twocolumn,
tightenlines]{revtex4}%
\usepackage{graphicx}
\usepackage{amsmath,bm}
\usepackage[mathscr]{euscript}
\usepackage{dcolumn}% Align table columns on decimal point

\begin{document}
\title{Emergence of a Kondo singlet state with the Kondo temperature well 
beyond 1,000K in the proton-embedded electron gas: 
Possible route to high-${\bm T}_{\bm c}$ superconductivity}

\author{Yasutami Takada}
\affiliation{Institute for Solid State Physics, University of Tokyo,
             Kashiwa, Chiba 277-8581, Japan}
%\address{Institute for Solid State Physics, University of Tokyo,
%         5-1-5 Kashiwanoha, Kashiwa, Chiba 277-8581, Japan}

\author{Ryo Maezono}
\affiliation{School of Information Science, JAIST, 
1-1 Asahidai, Nomi, Ishikawa 923-1292, Japan} 

\author{Kanako Yoshizawa} 
\affiliation{Institute for Solid State Physics, University of Tokyo,
%\address{Institute for Solid State Physics, University of Tokyo,
             Kashiwa, Chiba 277-8581, Japan}
 
%\date{\today}

%%%%%%%%%%%%%%%%%%%%%%%%%%%< Abstract: 236 words >%%%%%%%%%%%%%%%%%%%%%%%%%%%%%%
\begin{abstract}
Hydrogen in metals has attracted much attention for a long time from both basic 
scientific and technological points of view. Its electronic state has been 
investigated in terms of a proton embedded in the electron gas mostly by the 
local density approximation (LDA) to the density functional theory. At high 
electronic densities, it is well described by a bare proton H$^+$ screened by 
metallic electrons (charge resonance), while at low densities two electrons are 
localized at the proton site to form a closed-shell negative ion H$^-$ protected 
from surrounding metallic electrons by the Pauli exclusion principle. However, 
no details are known about the transition from H$^+$ to H$^-$ in the 
intermediate-density region. Here, by accurately determining the ground-state 
electron distribution $n({\bm r})$ by the use of LDA and diffusion Monte Carlo 
simulations with the total electron number up to 170, we obtain a complete 
picture of the transition, in particular, a sharp transition from short-range 
H$^+$ screening charge resonance to long-range Kondo-like spin-singlet 
resonance, the emergence of which is confirmed by the presence of an anomalous 
Friedel oscillation characteristic to the Kondo singlet state with the Kondo 
temperature $T_{\rm K}$ well beyond 1,000K. This study not only reveals 
interesting competition between charge and spin resonances, enriching the 
century-old paradigm of metallic screening to a point charge, but also 
discovers a long-sought high-$T_{\rm K}$ system, opening an unexpected route to 
room-temperature superconductivity in a Kondo lattice made of protons. 
\end{abstract}

%67.85.LM Degenerate Fermi gases
%71.10.Ca Electron gas, Fermi gas 
%71.15.Mb Density functional theory, 
%	local density approximation, gradient and other corrections 
%71.45.Gm Exchange, correlation, dielectric and magnetic response functions, 
%         plasmons 
%71.55.-i Impurity and defect levels 
%75.20.Hr Local moment in compounds and alloys; Kondo effect, 
%         valence fluctuations, heavy fermions 
%74.70.Tx Heavy-fermion superconductors (for heavy-fermion systems in 
%         magnetically ordered materials, see 75.30.Mb; see also 
%71.27.+a Strongly correlated electron systems, heavy fermions) 
%71.15.Pd Molecular dynamics calculations (Car-Parrinello) and other 
%         numerical simulations 

\maketitle

%\section{}
%\label{}

%%%%%%%%%%%%%%%%%%%%%%%%%%%%%%%%%[ Section 1 ]%%%%%%%%%%%%%%%%%%%%%%%%%%%%%%%%%%
\section{Introduction}
\label{Sec_1}

\begin{figure}[htbp]
\begin{center}
\includegraphics[scale=0.42,keepaspectratio]{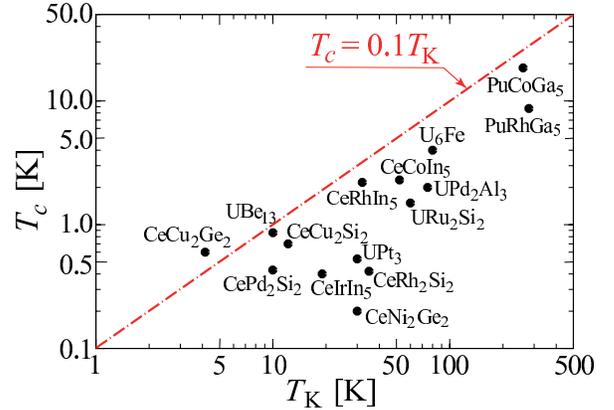}
\end{center}
\caption[Fig.1]{(Color online) Superconducting transition temperature $T_c$ 
versus Kondo temperature $T_{\rm K}$ (a characteristic energy scale for spin 
fluctuations) in heavy fermion superconductors.}
\label{fig:1}
\end{figure}

%%%%%%%%%%%%%%%%%< Paragraph 1: Heavy-Fermion Superconductors >%%%%%%%%%%%%%%%%%
Physics in the heavy fermion superconductors has been understood by the concept 
of quantum criticality in a system of regularly arrayed dense Kondo impurities 
(Kondo lattice)~\cite{Ref10,Ref10p,Ref10q} and the spin-fluctuation mechanism 
is believed to be responsible for superconductivity, as inferred from the 
strong correlation between the superconducting transition temperature $T_c$ 
and the Kondo temperature $T_{\rm K}$~\cite{Ref30p,Ref30q,Ref30,Ref30a}. More 
specifically, $T_c$ is of the order of $0.1 T_{\rm K}$, as shown in 
Fig.~\ref{fig:1} plotted based on the information available in the literature, 
from which we can conceive an idea that high-$T_c$ 
superconductivity will be obtained if we can discover a Kondo system with very 
high $T_{\rm K}$. In fact, the recently-discovered plutonium compounds such as 
PuCoGa$_5$ with $T_c=18.5$K and $T_{\rm K}\approx 260$K~\cite{Ref30r,Ref30s} may 
be regarded as the successful realization of this idea. Thus we should make 
further pursuit of this idea by searching for a new class of Kondo systems with 
$T_{\rm K}$ higher than 1,000K. Theoretically, this search can be done by the 
first-principles quantitative determination of $T_{\rm K}$ for the composite 
system of an impurity atom embedded in a metal.

%%%%%%%%%%%%%%%%%%%%%%%%%%%%%%%%%%< Figure 1 >%%%%%%%%%%%%%%%%%%%%%%%%%%%%%%%%%%
%------------------------------------------------------------------------------%

%%%%%%%%%%%%%%%%%< Paragraph 2: History of metallic screening >%%%%%%%%%%%%%%%%%
As first suggested by Debye and H\"uckel~\cite{Ref00}, an atomic nucleus charge 
$+Ze$ in a metal is screened by accumulation of metallic 
electrons which is regarded as a charge resonance and well described by the 
linear response theory. This concept of metallic screening is prevailing for 
a century, but because $Z$ is not infinitesimally small, we need to consider 
nonlinear effects in the screening, including the contribution from spin 
fluctuations. The spin contribution will be enhanced, if $Z$ is an odd 
integer, such as $Z=1$ (case of a proton), in which a spin-polarized bound 
state might appear at the impurity atom. 

%%%%%%%%%%%%%< Paragraph 3: History of hydrogen impurity in metals >%%%%%%%%%%%%
With the above basic scientific issues in mind, we have concerned with the problem 
of hydrogen impurity in metals which attracts long attention from a technological 
point of view~\cite{Ref0}, such as hydrogen storage in solids, sensor applications, 
and catalysis. Its electronic state has been investigated in terms of a proton 
immersed into an interacting many-electron system plus a compensating background 
(electron gas: EG) since 1970s, because this is an ideal system to study important 
topics related to an impurity in metals, such as the embedding energy, a key 
quantity in the effective-medium theory~\cite{Ref3,Ref4,Ref5}. It is also studied 
from a motivation to improve on the local density approximation (LDA) to the 
density functional theory (DFT) by using the electron distribution $n({\bm r})$ 
obtained by quantum Monte Carlo methods~\cite{Ref6,Ref7,Ref8}. Nonlinear metallic 
screening is another extensively examined topic in this 
system~\cite{Ref1,Ref2,Ref2a}, but no serious attention has been paid 
so far to the spin resonance effect. 

\begin{figure}[htbp]
\begin{center}
\includegraphics[scale=0.46,keepaspectratio]{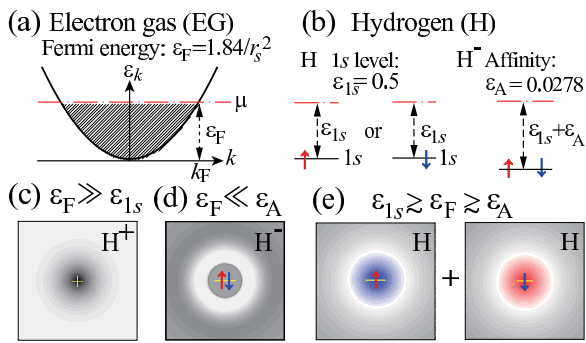}
\end{center}
\caption[Fig.2]{(Color online) Proton-embedded EG with characteristic energies 
in (a) and (b). Three possible ground states are schematically illustrated in 
(c)-(e) corresponding, respectively, to a bare proton H$^+$ screened by metallic 
electrons with the screening length $\approx k_{\rm F}^{-1}$ (a CR state), 
a closed-shell ion H$^-$ confined in EG, and an SSR state in which if the 1$s$ 
(virtual) level of H is temporarily occupied by a single up- (down-) spin 
electron, down- (up-) spin clouds of itinerant electrons are formed around H 
for spin screening with the screening length $\xi_{\rm K}\gg k_{\rm F}^{-1}$. 
These temporary states are superposed with interchanging the roles of spins 
to make a resonance state $\Psi_{\rm SSR}$.}
\label{fig:2}
\end{figure}

%%%%%%%%%%< Paragraph 4: Description of the system & on H^+ & H^- >%%%%%%%%%%%%%
The homogeneous EG with the average density $n_0$ is specified by a single 
parameter $r_s$, defined by $r_s\!=\!(3/4\pi n_0)^{1/3}$ in units of the Bohr 
radius $a_{\rm B}$. (We use atomic units hereafter.) Its characteristic energy 
is the Fermi energy $\varepsilon_{\rm F}$, given by $k_{\rm F}^2/2 \,(=\!1.84
r_s^{-2})$ with $k_{\rm F} (=\!1.92r_s^{-1})$ the Fermi momentum 
(Fig.~\ref{fig:2}(a)). Hydrogen, on the other hand, has two typical energies 
(Fig.~\ref{fig:2}(b)), the $1s$ level $\varepsilon_{1s}(=\!0.5)$ and the 
electron affinity $\varepsilon_{\rm A}(=\!0.0278)$. Ratio of $\varepsilon_{\rm F}$ 
to $\varepsilon_{1s}$ or $\varepsilon_{\rm A}$ determines the character of the 
ground state in the proton-embedded EG; for high $n_0$ corresponding to 
$\varepsilon_{\rm F}\!\gg \!\varepsilon_{1s}$ (or $r_s \!\ll\!2$), the $1s$ level 
is buried in the continuum of EG and thus no electrons are bound to a bare 
proton H$^+$, leading to a charge resonance (CR) state in which H$^+$ is 
dielectrically screened by accumulation of itinerant electrons near the Fermi 
level $\mu$ (Fig.~\ref{fig:2}(c)). For low $n_0$ with $\varepsilon_{\rm F}\!\ll 
\!\varepsilon_{\rm A}$ (or $r_s \!\gg\!10$), H$^+$ captures two antiparallel-spin 
electrons to form H$^-$. This closed-shell negative ion resides in EG with 
repelling other electrons owing to the Pauli exclusion principle 
(Fig.~\ref{fig:2}(d)), but if $\varepsilon_{\rm F}$ increases and reaches 
as high as $\varepsilon_{\rm A}$, the Fermi pressure from EG to the ion becomes 
so large that the localized electrons in H$^-$ begin to spill out into EG. 
Then a crucial question is whether this state at $\varepsilon_{\rm F} \!\approx 
\!\varepsilon_{\rm A}$ is the same as that in Fig.~\ref{fig:2}(c) or not.  

%%%%%%%%%%%%%%%%%%%%%%%%%%%%%%%%%%< Figure 2 >%%%%%%%%%%%%%%%%%%%%%%%%%%%%%%%%%%
%------------------------------------------------------------------------------%

%%%%%%%%%%%%< Paragraph 5: Discussion on Kondo-like resonance state >%%%%%%%%%%%
Intuitively, for $\varepsilon_{\rm A}\!\lesssim \!\varepsilon_{\rm F}\! 
\lesssim \!\varepsilon_{1s}$, we can imagine a spin-polarized state made of a 
single electron with either up or down spin captured by H$^+$, but in view of 
the concept of spin screening to form a Kondo singlet~\cite{Ref10,Ref10a} in 
the impurity Anderson model (IAM)~\cite{Ref11}, we anticipate the emergence of 
not a spin-polarized but a Kondo-like spin-singlet resonance (SSR) state 
$\Psi_{\rm SSR}$ (Fig.~\ref{fig:2}(e)). Because there is no clear distinction 
between conduction and localized electrons, this SSR state is composed of only 
itinerant electrons near $\mu$ without a local spin moment, similar to CR, but 
an important difference exists in the screening length; for CR, it is the 
Thomas-Fermi length $\lambda_{\rm TF}\!\approx \!k_{\rm F}^{-1}$, but for SSR, 
the Kondo-screening length $\xi_{\rm K}$ is much longer than $k_{\rm F}^{-1}$, 
leading to an anomalous Friedel oscillation~\cite{Ref12a,Ref12,Ref12b,Ref12c}. 
Then the main aim of this paper is to confirm this conjecture about the 
emergence of SSR in the proton-embedded electron gas with determining 
$T_{\rm K}$ from first principles, but this confirmation is not an easy task 
due to the existence of various difficulties, as we shall explain below 
in some detail. 

%%%%%%%%%%%%%%%%%%%%< Paragraph 6: Impurity Anderson Model >%%%%%%%%%%%%%%%%%%%%
The Hamiltonian for IAM, $H_{\rm A}$, is written as~\cite{Ref11}
\begin{align}
H_{\rm A}=&\sum_{{\bm k}\sigma}\varepsilon_{\bm k}
c^{\dag}_{{\bm k}\sigma}c_{{\bm k}\sigma}
+E_d\sum_{\sigma}c^{\dag}_{d\sigma}c_{d\sigma}
+Uc^{\dag}_{d\uparrow}c_{d\uparrow}c^{\dag}_{d\downarrow}c_{d\downarrow}
\nonumber \\
&+\sum_{{\bm k}\sigma}V_{d{\bm k}}(c^{\dag}_{d\sigma}c_{{\bm k}\sigma}
+c^{\dag}_{{\bm k}\sigma}c_{d\sigma}),
\label{eq:0}
\end{align}
in second quantization with use of the annihilation operator $c_{{\bm k}\sigma}$ 
for a conduction electron with wave vector ${\bm k}$, spin $\sigma$ and 
one-body band energy $\varepsilon_{\bm k}$, while $c_{d\sigma}$ is an 
operator to destroy a localized $d$ electron with spin $\sigma$ at the energy 
level $E_d$. The $d$ electrons not only interact to each other at the 
localized site with the strength $U$ but also hybridize with the conduction 
electrons with the strength $V_{d{\bm k}}$. Based on $H_{\rm A}$, 
$\Psi_{\rm SSR}$ is given as~\cite{Ref10b,Ref10c}
\begin{align}
\Psi_{\rm SSR}=a_0\Phi_0+\sum_{\bm k}a_{\bm k}
(c^{\dag}_{d\uparrow}c_{{\bm k}\uparrow}
+c^{\dag}_{d\downarrow}c_{{\bm k}\downarrow})\Phi_0,
\label{eq:0a}
\end{align}
where $\Phi_0$ is the Slater determinant made of conduction-electron orbitals 
and the parameters, $a_0$ and $a_{\bm k}$, are to be determined variationally. 
As Eq.~(\ref{eq:0a}) clearly shows, $\Psi_{\rm SSR}$ is not descibed by 
a single Slater determinant but is a correlated many-body state. It must also 
be noted that {\it continuum} conduction states around the Fermi level $\mu$ 
are indispensable for the construction of this $\Psi_{\rm SSR}$. 

%%%%%%%%%%%%%%%%%%%%%%< Paragraph 7: Atom embedded in EG >%%%%%%%%%%%%%%%%%%%%%%
In our work, we are not allowed to employ this simple model $H_{\rm A}$. 
Instead, we have to start with the first-principles Hamiltonian $H$, described 
in atomic units as 
\begin{align}
H \!=\! -\sum_i \! \frac{\mbox{\boldmath$\nabla$}_i^2}{2} \!+\! \frac{1}{2} 
\sum_{i \neq j} \frac{1}{|{\bm{r}_{i}}\! -\! {\bm{r}_{j}}|}  
\!+\! \sum_{i} v_{\rm ext}({\bm{r}_{i}})\!+\!C_{N}^Z,
\label{eq:1}
\end{align}
in first quantization. Here we have considered a neutral atom of atomic 
number $Z$ at the origin of coordinates immersed into the jellium sphere of 
radius $R$ and average density $n_0$. The number of electrons contained in 
the jellium sphere is $4\pi R^3n_0/3\!=\!(R/r_s)^3$, so that the total electron 
number $N$ is equal to $Z\!+\!(R/r_s)^3$, satisfying global neutrality, 
from which we obtain $R=(N\!-\!Z)^{1/3}r_s$. In Eq.~(\ref{eq:1}), the external 
potential working on an electron $v_{\rm ext}({\bm r})$ is composed of the 
potential from the nucleus and that from the positive background, written as
\begin{align}
v_{\rm ext}({\bm r})\!=\!& -\frac{Z}{|{\bm r}|}
-\frac{N-Z}{2}\frac{3R^2-{\bm r}^2}{R^3}\theta(R-|{\bm r}|)
\nonumber \\
&-\frac{N-Z}{|{\bm r}|}\theta(|{\bm r}|-R),
\label{eq:2}
\end{align}
with $\theta(x)$ the Heaviside function and $C_{N}^Z$ represents 
the Coulomb self-energy stemming from both nucleus-background and 
intra-background interactions, given by 
\begin{align}
C_{N}^Z \!=\! \frac{3}{2}\frac{Z(N-Z)^{2/3}}{r_s}
+\frac{3}{5}\frac{(N-Z)^{5/3}}{r_s}.
\label{eq:3}
\end{align}
In solving Eq.~(\ref{eq:1}), we impose the fixed boundary condition to make the 
wave function vanish at $|{\bm r}_i|\!=\!R$. Irrespective of whether we include 
the constant term $C_{N}^Z$ in $H$ or not, there is no problem of divergence 
in this finite-$N$ system, but $C_{N}^Z$ is needed in order 
to achieve the mutual global cancellation in energies between the 
electron-background attractive potentials and the repulsive potentials of both 
electron-electron and intra-background interactions for the bulk ($N \! \to \! 
\infty$) system~\cite{Ref14}. 

%%%%%%%%%%%%%%%%%< Paragraph 8: Comparison between H_A and H >%%%%%%%%%%%%%%%%%%
By comparing $H$ in Eq.~(\ref{eq:1}) with $H_{\rm A}$, we readily see their 
differences; first, there is no predetermined localized state in $H$ and thus 
a (virtual) localized state, if any, must be determined in the first place in 
constructing $\Psi_{\rm SSR}$ with related parameters such as $E_d$, $U$, and 
$V_{d{\bm k}}$, if necessary. Second and more importantly, the long-range 
Coulomb interaction, which is completely neglected in $H_{\rm A}$, works 
among metallic electrons in $H$, making solution of the problem quite difficult. 
In fact, even in the homogeneous EG without the embedded atom, the problem is 
sufficiently complicated in both variational~\cite{Ref14a,Ref14b,Ref14c} and 
Green's-function~\cite{Ref14d,Ref14e} approaches. Incidentally this long-range 
Coulomb interaction is the source to bring about CR and therefore it is 
indispensable for discussing competition between CR and SSR. Since this 
discussion constitutes another important aim of this paper, we can never 
neglect this long-range Coulomb interaction, making all theoretical and 
computational techniques developed so far for $H_{\rm A}$ useless to $H$. 

%%%%%%%%%%%%%%%%%%< Paragraph 9: Difficulties of DMC and DFT >%%%%%%%%%%%%%%%%%%
Usually, the first-principles Hamiltonian is solved by either diffusion Monte 
Caro (DMC) simulations or DFT-based methods. The former is an excellent method 
to obtain fairly accurate results for the ground state, but it can never 
directly treat $\Psi_{\rm SSR}$, because DMC simulations can be done only for 
finite-$N$ systems in which all levels are discrete, while in constructing 
$\Psi_{\rm SSR}$, we need continuum conduction states which are allowed only in 
the bulk ($N\! \to \! \infty$) system. The latter methods can easily treat the 
bulk system, but the ground-state physical quantities are calculated in terms 
of a single Slater determinant made of Kohn-Sham (KS) orbitals introduced in 
DFT, so that it is not clear at all as to how much the obtained quantities 
reflect the highly correlated many-body nature of $\Psi_{\rm SSR}$ and how 
accurate they are, especially because in actual calculations we always have to 
resort to some approximation to the exchange-correlation energy functional 
$E^{\rm xc}[n({\bm r})]$ such as LDA. 

%%%%%%%%%%%< Paragraph 10: Basic strategy: n(r) as a basic quantity  >%%%%%%%%%%
Faced with those difficulties, we have decided to focus on $n({\bm r})$ rather 
than the wave function $\Psi_{\rm SSR}$ itself, mainly because DFT can, 
in principle, provide {\it exact} $n({\bm r})$ and the corresponding 
ground-state energy $E_0$ by projecting the real interacting many-body system 
to a fictitious auxiliary non-interacting system in which $n({\bm r})$ can be 
calculated with use of a single Slater determinant made of KS orbitals, even if 
we know nothing about $\Psi_{\rm SSR}$ in the real system. This nontrivial 
assertion, one of the central theorems in DFT, is {\it rigorously proved} as 
long as the ground state is non-degenerate~\cite{Ref10d}, as is the case for SSR. 
Of course, information obtained only through $n({\bm r})$ and $E_0$ is limited 
and useless for discussing transport~\cite{Ref10e} and excited-state properties, 
but we claim that it is still plenty enough for our purpose of distinguishing 
between CR and SSR states and determining $T_{\rm K}$ in SSR. In the context of 
DFT, given exact $n({\bm r})$, it is also an interesting issue to clarify how KS 
orbitals, which are defined in the fictitious system and thus have basically no 
physical relevance, behave so as to correctly provide $n({\bm r})$ in a 
strongly-correlated state such as SSR. Thus this clarification constitutes an 
additional aim of this paper. 

%%%%%%%%%%%%%%%%%%%< Paragraph 11: Implementation procedure  >%%%%%%%%%%%%%%%%%%
In implementing calculations of $n({\bm r})$, we adopt the following 
strategies: (i) In actual LDA calculations, we employ the local spin-density 
approximation (LSDA)~\cite{RefA1} by choosing a spin-dependent 
exchange-correlation energy functional $E^{\rm xc}[n_{\sigma}({\bm r})]$ written 
in terms of the spin-resolved electron distribution $n_{\sigma}({\bm r})$, 
so that we can check a possibility of the spin-polarized (i.e., 
non-spin-singlet) ground state by detecting the difference between 
$n_{\uparrow}({\bm r})$ and $n_{\downarrow}({\bm r})$. (ii) We assess the 
results in LSDA at finite-$N$ systems in comparison with those in fixed-node 
DMC~\cite{Ref13} with taking $N$ up to 170, a much larger size than those in 
previous variational Monte Carlo (VMC) calculations~\cite{Ref7,Ref8}. 
(iii) We obtain $n({\bm r})$ in the bulk system by the calculation in LSDA and 
also by extrapolation of DMC data to $N\! \to \!\infty$. The extrapolated 
results for $n({\bm r})$ in DMC are independent of $N$ and thus we can assume 
that they will be free from any restrictions incurred from the fixed-node 
approximation, the only approximation adopted in DMC simulations, because the 
fixed-node positions in DMC are prescribed by $R$ (and consequently by $N$ for 
given $r_s$ due to $R\!=\!(N\!-\!Z)^{1/3}r_s$) in the fixed boundary condition, 
indicating that independence of $N$ also suggests independence of the postulated 
node positions. (iv) We check whether the obtained $n({\bm r})$ at $N\! \to \! 
\infty$ exhibits the behavior characteristic to the Kondo SSR state or not. 
More specifically, we look for {\it modulation of the Friedel-oscillation 
period}, a very important inherent property of the anomalous Friedel oscillation, 
to which we call serious attention for the first time in quantitatively 
determining $\xi_{\rm K}$ and consequently $T_{\rm K}$ from first principles.

%%%%%%%%%%%%%%%%%%< Paragraph 12: Summary of our results 1 >%%%%%%%%%%%%%%%%%%%%
In accordance with those strategies, we have investigated $n({\bm r})$ 
in finite-$N$ systems in both LSDA and DMC to find spin-unpolarized ground 
states with a strong size effect and a series of {\it magic numbers} (10, 60, 
170, $\cdots$) of $N$ at which convergent results are easily obtained. These 
features can be explained in terms of the emergence of SSR with its 
long-range nature of $\xi_{\rm K}$. Its emergence is also signaled in LSDA at 
$N\! \to\! \infty$ for $r_s\!\gtrsim\! 2$ by the appearance of a strange 
shallow bound KS orbital below the conduction band bottom with an unusually long 
binding radius. In spite of the large size effect, for {\it each} $N$, 
$n({\bm r})$ in LSDA agrees very well with that in DMC. 

%%%%%%%%%%%%%%%%%%%%%%%%%%%%%%%%%%< Figure 3 >%%%%%%%%%%%%%%%%%%%%%%%%%%%%%%%%%%
\begin{figure}[htbp]
\begin{center}
\includegraphics[scale=0.44,keepaspectratio]{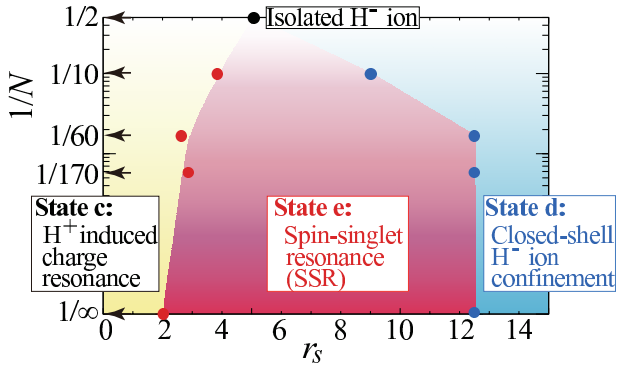}
\end{center}
\caption[Fig.3]{(Color online) Ground-state diagram in the proton-embedded 
electron gas in ($r_s$,$N^{-1}$) space with $N$ the total electron number, 
indicating sharp but size-dependent sequential transitions among CR, SSR, and 
closed-shell ion H$^-$. Rigorously speaking, SSR is defined only at $N \! \to \! 
\infty$, but the states for finite $N$ directly connected to SSR at $N \! \to \! 
\infty$ are also called SSR. In the bulk system, the screening length 
$\xi_{\rm K}$ is predicted to diverge at the CR-SSR boundary or $r_s\approx 1.97$ 
with the change of $r_s$, signaling the sharp transition. For an isolated H$^-$ 
ion at $N\!=\!2$, we have used the exact data for $n({\bm r})$~\cite{Ref15} 
to determine $r_s\!=\!5.04$ by averaging the local $r_s({\bm r}) (=\![3/4\pi 
n({\bm r})]^{1/3})$ over the weight of  $n({\bm r})$ itself. 
}
\label{fig:3}
\end{figure}
%------------------------------------------------------------------------------%

%%%%%%%%%%%%%%%%%%< Paragraph 13: Summary of our results 2 >%%%%%%%%%%%%%%%%%%%%
By summarizing the results thus calculated, we have obtained a ground-state 
diagram in ($r_s$,$N^{-1}$) space, shown in Fig.~\ref{fig:3}, exhibiting sharp 
sequential transitions among CR, SSR, and closed-shell H$^-$ ion confinement 
states. Contrary to the previous explanation~\cite{Ref3}, we claim that the 
very shallow bound KS orbital found in LSDA at intermediate densities is not a 
physical H$^-$ but appears just to describe the long-range change of 
$n(\bm{r})$ over $\xi_{\rm K}$ in SSR in the form of an {\it envelope density}. 
We also find that hydrogen is most stably embedded in EG in the SSR region, 
especially, optimally firmly at $r_s\! \approx\! 4$ with the Kondo temperature 
$T_{\rm K}\! \approx $2,100K, indicating our success in discovering a 
long-sought high-$T_{\rm K}$ system. Thus we may expect that 
superconductivity occurs at a temperature as high as about $0.1T_{\rm K}$ in a 
metallic hydrogen alloy at ambient pressure in which a macroscopic number of 
protons are regularly embedded in a metal in this density region to form a 
Kondo lattice. 

%%%%%%%%%%%%%%%%%%%< Paragraph 14: Contents of this paper >%%%%%%%%%%%%%%%%%%%%%
In Sec.~II, we explain the calculation methods in both LSDA and DMC. In Sec.~III, 
we show the calculated results and in Sec.~IV we discuss on the obtained results, 
together with their implications and future directions. Finally in Sec.~V, we 
give a summary of this paper. 

%%%%%%%%%%%%%%%%%%%%%%%%%%%%%%%%%[ Section 2 ]%%%%%%%%%%%%%%%%%%%%%%%%%%%%%%%%%%
\section{Calculation Methods}

%%%%%%%%%%%%%%%%%%%< Paragraph 15: LSDA for finite system >%%%%%%%%%%%%%%%%%%%%%
\subsection{LSDA in the finite-$N$ system}
\label{Subsec:2A}

Let us consider the neutral system of a single nucleus with atomic number $Z$ 
embedded in the $N$-electron jellium sphere of radius $R$. Its Hamiltonian $H$ 
is given in Eq.~(\ref{eq:1}). In LSDA to DFT, the KS equation is written as
\begin{equation}
\left [ -\mbox{\boldmath$\nabla$}^2/2 + v^{\rm KS}_{\sigma}(\bm{r}) 
\right ] \phi_{i\sigma} (\bm{r}) = \varepsilon_{i\sigma}\phi_{i\sigma}(\bm{r}),
\label{eq:4}
\end{equation}
where $\varepsilon_{i\sigma}$ and $\phi_{i\sigma}$ are the energy level and 
the normalized wave function for KS orbital $i$ and spin $\sigma$, respectively, 
and $v^{\rm KS}_{\sigma}(\bm{r})$ is the KS potential, determined by 
\begin{align}
v^{\rm KS}_{\sigma}(\bm{r})\! =\! v_{\rm ext}({\bm r})\!+\!
\int \! d\bm{r}'\, \frac{n(\bm{r}')}{|\bm{r}\! - \!\bm{r}'|}
\!+\!v^{\rm xc}_{\sigma} ({\bm r};[n_{\sigma}]),
\label{eq:5}
\end{align}
where $v^{\rm xc}_{\sigma} (\bm{r};[n_{\sigma}])$ is derived from 
$E^{\rm xc}[n_{\sigma}]$ through the functional derivative as
\begin{equation}
v^{\rm xc}_{\sigma} (\bm{r};[n_{\sigma}])= 
\delta E^{\rm xc}[n_{\sigma}]/\delta n_{\sigma}(\bm{r}).
\label{eq:6}
\end{equation}
With use of the lowest-$N_{\sigma}$ KS orbitals, $n_{\sigma}(\bm{r})$ is given by 
\begin{equation}
n_{\sigma}(\bm{r}) = \sum_{i=1}^{N_{\sigma}} |\phi_{i\sigma} (\bm{r})|^2, 
\label{eq:7}
\end{equation}
and $n(\bm{r})$ is the sum of $n_{\uparrow}(\bm{r})$ and 
$n_{\downarrow}(\bm{r})$. The spin density $n_{\sigma}(\bm{r})$ and 
consequently $N_{\sigma}$ with $N=\sum_{\sigma}N_{\sigma}$ should be determined 
by the self-consistent solution of Eqs.~(\ref{eq:4})-(\ref{eq:7}), together 
with the fixed boundary condition 
\begin{equation}
\phi_{i\sigma} (\bm{r})=0,
\label{eq:7a}
\end{equation}
at $|{\bm r}|\!=R\!=\!(N\!-\!Z)^{1/3}r_s$. By using those converged quantities, 
we can calculate $E_0(N,Z)$ the ground-state energy including the constant 
term $C_{N}^Z$ by
\begin{align}
E_0(N,Z)\!=&\!\sum_{i\sigma}\varepsilon_{i\sigma}\!+\!\sum_{\sigma}\!
\int \! d\bm{r}[v_{\rm ext}({\bm r})\!-\! v^{\rm KS}_{\sigma}(\bm{r})]
n_{\sigma}(\bm{r})
\nonumber \\
+&\frac{1}{2}\!\int \! \int \! d\bm{r}d\bm{r}'
\frac{n(\bm{r})n(\bm{r}')}{|\bm{r}\! -\! \bm{r}'|}\!+\!E^{\rm xc}[n_{\sigma}]
\!+\!C_{N}^Z.
\label{eq:8}
\end{align}

%%%%%%%%%%%%%%%%%%%%%%%%%< Paragraph 16: VMC scheme >%%%%%%%%%%%%%%%%%%%%%%%%%%%
\subsection{VMC}
\label{Subsec:2B}

With use of the lowest-$N$ KS orbitals thus obtained, we can define the Slater 
determinant $\Phi_0 ({\bm r}_1,\!\cdots\!,{\bm r}_N)$, with which the trial 
many-body ground-state wave function $\Phi({\bm r}_1,\!\cdots\!,{\bm r}_N)$ 
for the VMC calculation can be constructed in the Slater-Jastrow type 
as~\cite{Ref17}
\begin{align}
\Phi({\bm r}_1,\!\cdots\!,{\bm r}_N)\! = \!
\exp [J({\bm r}_1,\!\cdots\!,{\bm r}_N)]
\Phi_0 ({\bm r}_1,\!\cdots\!,{\bm r}_N),
\label{eq:9}
\end{align}
where the Jastrow function $J({\bm r}_1,\!\cdots\!,{\bm r}_N)$ contains the 
terms to describe electron-nucleus correlation $u_1({\bm r}_i)$, 
two-electron correlation $u_2({\bm r}_i\!-\!{\bm r}_j)$, 
and three-body nucleus-two-electron correlation 
$u_3({\bm r}_i,{\bm r}_j,{\bm r}_i\!-\!{\bm r}_j)$ as
\begin{align}
J({\bm r}_1,\!\cdots\!,{\bm r}_N)
=& \sum_i u_1({\bm r}_i)+\sum_{i>j}u_2({\bm r}_i\!-\!{\bm r}_j)
\nonumber \\
&+\sum_{i>j} u_3({\bm r}_i,{\bm r}_j,{\bm r}_i\!-\!{\bm r}_j). 
\label{eq:9a}
\end{align}
The actual choice of the forms for $u_1({\bm r}_i)$, 
$u_2({\bm r}_i\!-\!{\bm r}_j)$, and $u_3({\bm r}_i,{\bm r}_j,
{\bm r}_i\!-\!{\bm r}_j)$ as well as their optimization is done by adopting 
the CHAMP-code package~\cite{Ref16} as it is. Then the expectation value 
$\langle A \rangle$ of an operator $A$ is given by
\begin{align}
\langle A \rangle^{\rm VMC} =\langle \Phi|A|\Phi \rangle/
\langle \Phi|\Phi \rangle.
\label{eq:10a}
\end{align}
By putting $A=\hat{n}({\bm r})=\sum_i\delta({\bm r}-{\bm r}_i)$ in 
Eq.~(\ref{eq:10a}), we obtain $n({\bm r})$ in VMC. 

%%%%%%%%%%%%%%%%%%%%%%%%%< Paragraph 17: DMC scheme >%%%%%%%%%%%%%%%%%%%%%%%%%%%
\subsection{DMC}
\label{Subsec:2C}

Starting with the variationally optimized wave function $\Phi$ thus determined, 
we can further improve on the ground-state wave function by considering the 
diffusion equation for $\Psi(\tau)$ in the imaginary time $\tau$ as
\begin{align}
-\frac{\partial\Psi(\tau)}{\partial \tau}=(H-E_R)\Psi(\tau),
\label{eq:9b}
\end{align}
where $E_R$ is the reference energy to be adjusted to $E_0$ in the course of 
DMC simulations by removing the $\tau$-dependence from the asymptotic form of 
$\Psi(\tau)$ at $\tau \! \to \! \infty$. Note that the formal solution to 
Eq.~(\ref{eq:9b}) is written as
\begin{align}
\Psi(\tau)=\sum_ne^{-(E_n-E_R)\tau}|\Psi_n\rangle \langle \Psi_n|\Phi\rangle,
\label{eq:9c}
\end{align}
where $\{\Psi_n\}$ is the normalized mutually-orthogonal complete set of eigen 
functions for $H$ with the corresponding set of eigen energies $\{E_n\}$. Then, 
as long as $\langle \Psi_0|\Phi\rangle \neq 0$, the asymptotic $\tau$-independent 
wave function $\Psi$ is reduced to the true ground-state wave function $\Psi_0$, 
apart from the normalization factor. 

%%%%%%%%%%%%%%%%< Paragraph 18: Comment on the Anderson theorem >%%%%%%%%%%%%%%%
It is appropriate to add a comment on the condition of $\langle \Psi_0|\Phi\rangle 
\neq 0$ here; by invoking the Anderson's orthogonality theorem~\cite{Ref16a}, 
one may argue that $\langle \Psi_0|\Phi\rangle$ vanishes in SSR, but this is 
not correct for the reasons below; (i) both $\Psi_0$ and $\Phi$ include the 
effect of the impurity atom, while the Anderson's theorem concerns with the 
relation between the wave functions with and without the impurity. (ii) Due 
to the presence of the Jastrow factor $J$, $\Phi$ is not simply given by the 
single Slater determinant $\Phi_0$ on which the Anderson's theorem is proved. 
(iii) DMC simulations are done for finite $N$, while the Anderson's theorem 
becomes valid only at $N \! \to \! \infty$. 

%%%%%%%%%%%%< Paragraph 19: Comment on the fixed-node approximation >%%%%%%%%%%%
In order to avoid the notorious fermion sign problem, we employ the fixed-node 
approximation in DMC simulations. This approximation may bring about undesirable 
errors in $\Psi$, but we try to minimize them by seeking for $N$-independent 
results by exploiting the fact that the node positions depend on $N$ in the 
fixed boundary condition, leading to the hope that unphysical node-position 
dependent effects will be removed by extracting the $N$-independent results. 
In performing actual fixed-node DMC simulations at a fixed $N$, we adopt CHAMP 
again to obtain the stably converged asymptotic wave function $\Psi({\bm r}_1,
\!\cdots\!,{\bm r}_N)$. Then $n({\bm r})$ in DMC is estimated by a second-order 
approximation to the exact expectation value~\cite{Ref17}, which amounts to
\begin{align}
n({\bm r})^{\rm DMC} = 2\,\langle \Psi| \hat{n}({\bm r}) |\Phi \rangle 
/\langle \Psi | \Phi \rangle - n({\bm r})^{\rm VMC},
\label{eq:10}
\end{align}
where $n({\bm r})^{\rm VMC}=\langle \hat{n}({\bm r}) \rangle^{\rm VMC}$. 

%%%%%%%%%%%%%%%%%%%< Paragraph 20: LSDA for the bulk system >%%%%%%%%%%%%%%%%%%%
\subsection{LSDA in the bulk system}
\label{Subsec:2D}

Contrary to VMC and DMC, LSDA allows us to directly treat the bulk ($N \! 
\to \! \infty$) system, in which almost all states in KS orbitals are continuum 
ones, for which we may write $\varepsilon_i\!=\!\bm{k}^2/2$ with 
momentum $\bm{k}$ and $\phi_{i\sigma}(\bm{r})\!=\!R_{kl\sigma}(r)
Y_{lm}(\bm{r}/r)$ with the spherical harmonics $Y_{lm}(\bm{r}/r)$ and the 
radial wave function $R_{kl\sigma}(r)$ satisfying the following boundary 
condition at $r\, (\equiv\! |\bm{r}|) \! \to \! \infty\,$: 
\begin{align}
R_{kl\sigma}(r) \to \cos[\delta_{l\sigma}(k)]j_l(kr)
-\sin[\delta_{l\sigma}(k)]n_l(kr),
\label{eq:12}
\end{align}
apart from a normalization factor, where $j_l(kr)$ and $n_l(kr)$ are the 
spherical Bessel functions and $\delta_{l\sigma}(k)$ is the phase shift of 
angular momentum $l$ to be determined under the condition of 
$\delta_{l\sigma}(\infty)\!=\!0$, ensuring that an electron behaves as a free 
particle at $k\, (\equiv\! |\bm{k}|)\! \to \!\infty$. There is a possibility 
of finding bound states below the bottom of the conduction band 
($\varepsilon_{i\sigma}\!<\!0$) among KS orbitals for which 
$\phi_{i\sigma}(\bm{r})\! \to \! 0$ at $r \! \to \! \infty$. The Levinson 
theorem~\cite{Ref18} dictates that the total number of the bound states in KS 
orbitals $N_{\rm BS}$ is given by $\sum_{l\sigma}\!(2l\!+\!1)\delta_{l\sigma}(0)
/\pi$, while $Z\!=\!\sum_{l\sigma}(2l\!+\!1)\delta_{l\sigma}(k_{\rm F})/\pi$ 
by the Friedel sum rule~\cite{Ref19}. 

%%%%%%%%%%%%%%%%%%%%%%%< Paragraph 21: Embedding energy >%%%%%%%%%%%%%%%%%%%%%%%
\subsection{Embedding energy}
\label{Subsec:2E}

The embedding energy $\delta E$ is defined as the difference of the 
ground-state energies between the atom-embedded EG and the 
system of separated homogeneous EG and neutral atom. Usually this concept is 
relevant only to the bulk EG. Thus, denoting the ground-state energy of the 
isolated neutral atom as $E_{\rm a}^Z$, we can obtain $\delta E$ through 
\begin{align}
\delta E=\lim_{N \to \infty} \left [E_0(N,Z)-E_0(N-Z,0)\right ]-E_{\rm a}^Z.
\label{eq:11}
\end{align}
Since $\delta E$ is of the order $O(1)$ in comparison with $E_0$ of 
the order $O(N)$, due care must be exerted in order to accurately 
evaluate $\delta E$ at $N \! \to \! \infty$. For this purpose, we rewrite 
Eq.~(\ref{eq:11}) with using $\delta n(\bm{r})\ [=\!n(\bm{r})-n_0]$ and 
$\delta_{l\sigma}'(k)$, the derivative of $\delta_{l\sigma}(k)$ with respect 
to $k$, into the following form with ensuring the convergence of integrals: 
\begin{align}
\delta E\!=&\!\sum_{i\sigma \in {\rm BS}} \varepsilon_{i\sigma}\!+\!
\sum_{l\sigma} (2l\!+\!1)\! \int\limits_0^{k_{\rm F}}\! dk \, 
 \,\frac{\delta_{l\sigma}'(k)}{\pi} \frac{k^2}{2}\notag \\
\!&-\!\int \! d{\bm r}\, \frac{Z}{r}\,\delta n(\bm{r})
\!+\!\frac{1}{2}\! \int \! \int \! d{\bm r} d{\bm r'} 
\frac{\delta n({\bm r})\delta n({\bm r'})}{|{\bm r}\!-\!{\bm r'}|}
\notag \\
\!&-\!\sum_{\sigma}\!\int \! d{\bm r} \left\{
v_{\sigma}^{xc}({\bm r};[n_{\sigma}])n_\sigma({\bm r})
-v_{\sigma}^{xc}({\bm r};[n_0/2])\frac{n_0}{2} \right\}
\notag \\
&+E^{xc}[n_\sigma]-E^{xc}[n_0/2]-E_{\rm a}^Z,
\label{eq:13}
\end{align}
where BS stands for the set of possible bound states. The difference 
in the contribution of KS energies from continuum states is treated by the 
consideration of change in the density of states, $\delta_{l\sigma}'(k)/\pi$. 

%%%%%%%%%%%%%%%%%%%%%%%%%%%%%%%%%[ Section 3 ]%%%%%%%%%%%%%%%%%%%%%%%%%%%%%%%%%%
\section{Results for Proton Immersion}

%%%%%%%%%%%%%%%%%%%< Paragraph 22: Results for bulk in LSDA >%%%%%%%%%%%%%%%%%%%
\subsection{LSDA for the bulk system}
\label{Subsec:3A}

\begin{figure}[htbp]
\begin{center}
\includegraphics[scale=0.50,keepaspectratio]{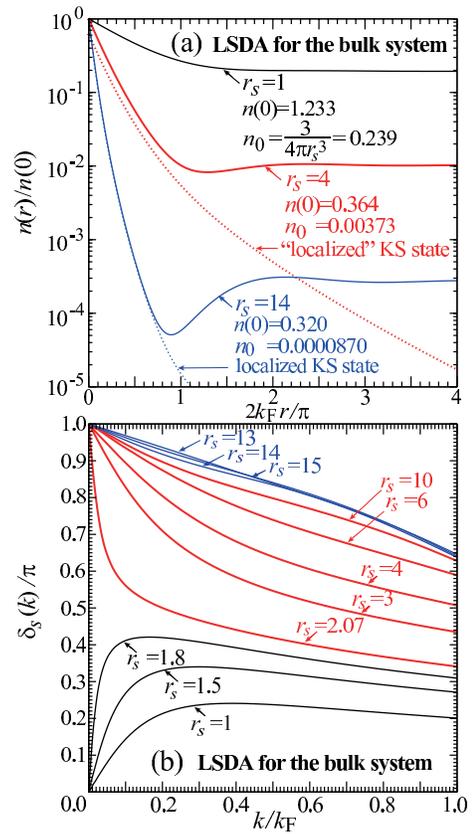}
\end{center}
\caption[Fig.4]{(Color online) Electron distribution $n(\bm{r})$ normalized by 
$n(0)$ in (a) and spin-independent s-wave phase shift $\delta_{s}(k)$ in (b) 
obtained in LSDA for the proton-embedded bulk EG. In (a), the dotted curves 
show the contribution from the KS bound state which is absent at $r_s\!=\!1$, 
must be a real physical state representing H$^-$ at $r_s\!=\!14$, and appears 
only as mathematical convenience for describing the long-range decrease in 
amplitude of the Friedel oscillation in SSR at $r_s\!=\!4$. The label 
``localized'' with a quotation mark indicates this situation. 
} 
\label{fig:4}
\end{figure}

In line with the previous result~\cite{Ref20}, the ground state in the 
proton-embedded ($Z\!=\!1$) bulk EG is perfectly spin-unpolarized, i.e., 
$n_{\uparrow}({\bm r})\!= \! n_{\downarrow}({\bm r})$ at every ${\bm r}$, 
at least for $r_s\!<\!15$ and our results on $n(\bm{r})$ and $\delta E$ in LSDA 
are in good agreement with those in previous 
calculations~\cite{Ref1,Ref2,Ref3,Ref4,Ref5,Ref6,Ref6a}. 
In Fig.~\ref{fig:4}, the obtained $n(\bm{r})$ normalized by $n(0)$ is plotted 
as a function of $2k_{\rm F}r/\pi$ for $r_s\!=\!1$, 4, and 14, together with 
the $s$-wave phase shift $\delta_{s}(k)$ which is spin-independent and gives 
by far the largest contribution among all $\delta_{l\sigma}(k)$s. 

%%%%%%%%%%%%%%%%%%%%%%%%%%%%%%%%%%< Figure 4 >%%%%%%%%%%%%%%%%%%%%%%%%%%%%%%%%%%
%------------------------------------------------------------------------------%

%%%%%%%%%%%%< Paragraph 23: Explanation of results for bulk in LSDA >%%%%%%%%%%%
For $r_s\!<\!1.97$, we obtain $\delta_{l\sigma}(0)\!=\!0$ and thus $N_{\rm BS}$ 
is zero, leading to the typical $n(\bm{r})$ in CR with H$^+$ screened by 
metallic electrons in a short range. For $r_s\!\ge \!1.97$, on the other hand, 
$\delta_{s}(0)\!=\!\pi$ and thus $N_{\rm BS}\!=\!2$, seemingly implying the 
sudden appearance of $H^-$ ion confined in EG at $r_s\!=\!1.97$~\cite{Ref3}. 
This must be true, if $n_{\rm L}(\bm{r})$ the localized-electron distribution 
is about the same as $n(\bm{r})$ for $|\bm{r}|$ smaller than the $H^-$ ion 
range, as is the case for $r_s\!\gtrsim \!12.5$, in which $\delta_{s}(k)$ is 
distinctive and almost a universal function of 
$k/k_{\rm F}$, as seen in Fig.~\ref{fig:4}(b). Note that the deep dip in 
$n(\bm{r})$ just outside the ion region, as seen for $r_s\!=\!14$ in 
Fig.~\ref{fig:4}(a), is a typical electron profile describing the repulsive 
action of localized closed-shell electrons to exclude metallic electrons 
from the ion region by the Pauli exclusion principle. 

%%%%%%%%%%%%%< Paragraph 24: Spin-resonance state for bulk in LSDA >%%%%%%%%%%%%
For $1.97\! <\! r_s\!\lesssim\!12.5$, however, the ``localized'' electrons 
behave much differently; at $r_s\!=\!4$, for example, they extend long 
up to $|\bm{r}|\! \approx \!82$ and concomitantly $n_{\rm L}(0)$ is much smaller 
than $n(0)$, indicating deep and massive penetration of itinerant electrons 
into the proton site, but such penetration would never be allowed due to the 
Pauli exclusion principle, if the closed-shell $H^-$ ion were firmly constructed. 
Thus, by remembering that KS orbitals in DFT are not necessarily tied with real 
physical entities but just introduced for mathematical convenience to correctly 
reproduce $n(\bm{r})$, we can assume that this ``localized'' orbital does not 
represent a real localized state but just describes the long-range change of 
$n(\bm{r})$ in the form of an {\it envelope density} over $\xi_{\rm K}$ in SSR. 
Notice that $\delta_{s}(k_{\rm F})$ in this density region lies between 
$0.7\!\times \!(\pi/2)$ and $1.3\!\times \!(\pi/2)$, which means that 
$\delta_{s}(k_{\rm F})$ is close to $\pi/2$, a value expected in the Kondo 
resonance state in IAM~\cite{Ref11}. 

%%%%%%%%%%%%%%%< Paragraph 25: Comparison between LSDA and DMC >%%%%%%%%%%%%%%%%
\subsection{Comparison between LSDA and DMC}
\label{Subsec:3B}

\begin{figure}[htbp]
\begin{center}
\includegraphics[scale=0.68,keepaspectratio]{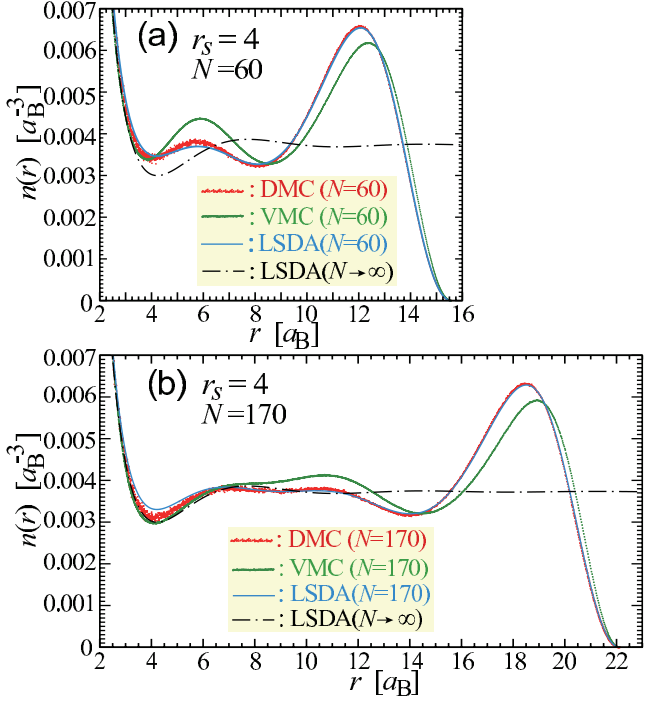}
\end{center}
\caption[Fig.5]{(Color online) Examples of the calculated electron distribution 
$n(\bm{r})$ obtained by DMC, VMC, and LSDA for (a) $N=60$ and (b) 170 at $r_s=4$. 
The result in LSDA at $N\!\to \!\infty$ is also shown.
} 
\label{fig:5}
\end{figure}

Before going into a more detailed discussion on SSR, let us assess the accuracy 
of LSDA in comparison with DMC, specifically at intermediate densities. At 
$r_s\!=\!4$, for example, in Fig.~\ref{fig:5}, we see a good agreement between 
LSDA and DMC for $n(\bm{r})$ at any $\bm {r}$, including the sphere boundary, 
at {\it each} $N$, though the results in VMC do not match so well, assuring 
the importance to perform DMC for taking the expectation values in 
accordance with Eq.~(\ref{eq:10}). 

%%%%%%%%%%%%%%%%%%%%%%%%%%%%%%%%%%< Figure 5 >%%%%%%%%%%%%%%%%%%%%%%%%%%%%%%%%%%
%------------------------------------------------------------------------------%

%%%%%%%%%%%%%%%%%%%%%%%%%< Paragraph 26: Size effect >%%%%%%%%%%%%%%%%%%%%%%%%%%
As for $N$ dependence or the size effect, we find that $N\!=\!170$ is not large 
enough to attain convergence in $n(\bm{r})$ for $|\bm{r}|\!\gtrsim \! 3.5$ 
in both LSDA and DMC. For smaller $|\bm{r}|$, however, no appreciable difference 
is seen between $N\!=\!60$ and 170 in DMC (and among $N\!=\!60$, 170, and 
$\infty$ in LSDA) for $r_s\! \gtrsim \!4$, implying that $N\!=\!60$ is large 
enough to obtain the convergent $n(\bm{r})$ near the proton site. 

%%%%%%%%%%%%%%< Paragraph 27: Cusp theorem and the on-top density >%%%%%%%%%%%%%
\subsection{Cusp theorem and the on-top density}
\label{Subsec:3C}

\begin{figure}[htbp]
\begin{center}
\includegraphics[scale=0.615,keepaspectratio]{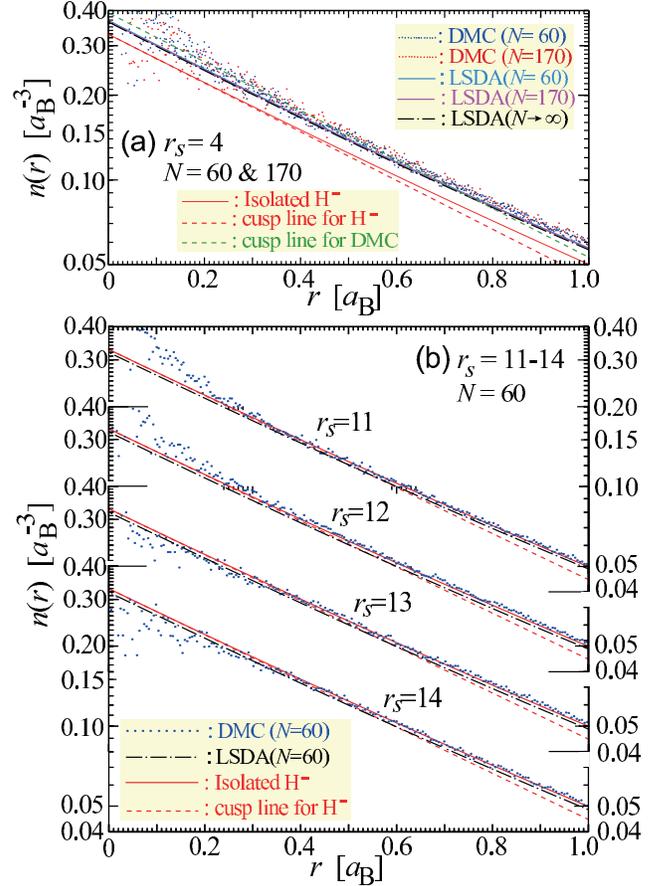}
\end{center}
\caption[Fig.6]{(Color online) Short-range part of $n(\bm{r})$ in DMC and LSDA 
at (a) $r_s=4$ and (b) $r_s=11-14$. For comparison, the exact result for an 
isolated negative hydrogen ion H$^-$~\cite{Ref15} is also plotted.
} 
\label{fig:6}
\end{figure}

According to the cusp theorem~\cite{Ref21}, $n(\bm{r})$ near a nucleus of 
atomic number $Z$ behaves rigorously in the manner as 
\begin{align}
n(\bm{r}) \xrightarrow[r\, \approx\, 0]{}n_{\rm cusp}(r)\equiv n(0)\exp(-2Zr),
\label{eq:14}
\end{align}
where $n_{\rm cusp}(r)$ exhibits strictly a linear change with $r$ in semilog 
plots (the dashed lines in Fig.~\ref{fig:6}(a) and (b)). By exploiting this 
linear behavior, we can rather easily and accurately determine the on-top 
density $n(0)$ in DMC by looking at the data in the region of $0.2 \!\lesssim
\!r\!\lesssim\!0.5$, in spite of the scattered nature of data points for 
$n(\bm{r})$ at $r\!\lesssim\!0.2$ due to the rapid increase in energy scale 
determined by the $r^{-1}$-Coulomb potential. 

%%%%%%%%%%%%%%%%%%%%%%%%%%%%%%%%%%< Figure 6 >%%%%%%%%%%%%%%%%%%%%%%%%%%%%%%%%%%
%------------------------------------------------------------------------------%

%%%%%%%%%%%%%%%%%%%%%%< Paragraph 28: n(0) for large r_s >%%%%%%%%%%%%%%%%%%%%%%
For $r_s\!\gtrsim \!11$, $n(0)$ thus obtained is about the same as the exact one 
for an isolated H$^-$ ion~\cite{Ref15}, as shown in Fig.~\ref{fig:6}(b) for 
$r_s=11-14$, indicating that the state is very close to the $H^-$ ion 
confinement state. A more detailed observation on the DMC data reveals a 
systematic change of the behavior at $r\!\lesssim\!0.2$ with the increase of 
$r_s$; for $r_s\!\lesssim\!12.5$, the majority of data points deviate upward, 
while opposite is the case for $r_s\!\gtrsim\!12.5$. The upward deviation 
indicates that the metallic electrons rather easily penetrate into the core 
of the H$^-$ ion, which is not allowed, once the closed-shell structure is 
solidly constructed. On the other hand, the downward deviation is consistent 
with the formation of the closed-shell ion, leading to the conclusion that the 
transition to the $H^-$ ion confinement state occurs at $r_s\!\approx\!12,5$, 
the same $r_s$ as that in LSDA. 

\begin{figure}[thbp]
\begin{center}
\includegraphics[scale=0.49,keepaspectratio]{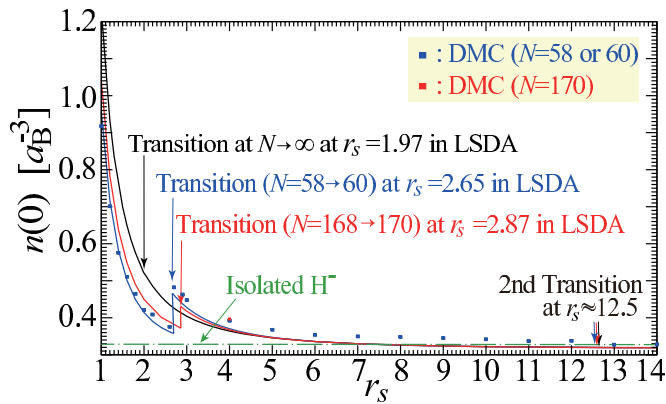}
\end{center}
\caption[Fig.7]{(Color online) On-top density $n(0)$ plotted as a function of 
$r_s$ for $N\!=\!60$ (and 170 only for $r_s=4$) in DMC and for $N\!=\!60$, 170, 
and $\infty$ in LSDA, together with the exact results for an isolated H$^-$ ion 
($N\!=\!2$). The transition points are indicated by arrows. 
For the data in DMC, errors are within the size of square symbols. 
} 
\label{fig:7}
\end{figure}

%%%%%%%%%%%%%%%%%%< Paragraph 29: n(0) as a function of r_s >%%%%%%%%%%%%%%%%%%%
In Fig.~\ref{fig:7}, we plot $n(0)$ by changing $r_s$ and $N$ and find that the 
results do not depend on $N$ for $r_s\!\gtrsim \!3$, but they do for $r_s\!
\lesssim\!3$, i.e., in the CR-SSR transition region. The transition is signaled 
by a jump in $n(0)$ in both LSDA and DMC, although its magnitude decreases with 
increasing $N$ and eventually at $N\!\to\!\infty$ no jump is seen in LSDA even 
at $r_s\!=\!1.97$ at which the transition is known to occur through the abrupt 
change in $\delta_{s}(0)$. 

%%%%%%%%%%%%%%%%%%%%%%%%%%%%%%%%%%< Figure 7 >%%%%%%%%%%%%%%%%%%%%%%%%%%%%%%%%%%
%------------------------------------------------------------------------------%

%%%%%%%%%%%%%%%%%%%%%< Paragraph 30: Physical meaning of N >%%%%%%%%%%%%%%%%%%%%
In an extensive search for favorable $N$ at which we can easily obtain the 
convergent ground state, we find that this jump becomes much enhanced at the 
magic numbers (10, 60, 170, $\cdots$) of $N$. Let us consider the reason for 
this fact; in LSDA, all KS levels are discrete at finite $N$ and the stacking 
sequences of the occupied levels are: $(1s,2p,2s)$, $(1s,2p,2s,3d,4f,5g,3p,3s)$, 
and $(1s,2p,2s,3d,4f,3p,3s,5g,4d,6h,7i,5f,8k,4p,4s)$ for $N=10$, 60, 170, 
respectively, reflecting the competition between the $-r^{-1}$ and $r^2$ 
potentials in Eq.~(\ref{eq:2}) and the outermost $s$ orbital, situating very 
near the Fermi level $\mu$ at each magic number $N$, plays a key role in 
stabilizing the ground state. We recognize that this outermost $s$ orbital 
mimics the SSR state at finite $N$, at least for $R<\xi_{\rm K}$, in view of 
the fact that the actual SSR state in Kondo physics is situated at $\mu$, 
extending very long over the range $\xi_{\rm K}$ with an $s$-wave character. 
Then the jump in $n(0)$ is related to the transition from the empty $s$-orbital 
state (corresponding to CR) to the $s$-orbital occupied one (corresponding to 
SSR) with changing $N$ in $8\! \to\! 10$, $58\! \to\! 60$ and $168\! \to\! 170$ 
at $r_s\!=\!3.85$, 2.65, and 2.87, respectively, in LSDA. In DMC, convergent 
results are also easily obtained for the same series of $N$ and the transition 
occurs, for example, at $r_s\!\approx\!2.65$ for $N\!=\!58\! \to \! 60$, just 
as in LSDA. The next $N$ in this series is 340 associated with the $5s$-orbital 
empty-occupied transition, but DMC at this $N$ is currently out of our reach. 
At the second transition into the H$^-$ confinement state, a change in the 
stacking sequence occurs in LSDA. Those transition points, along with the 
second transition point obtained in DMC by the data shown in Fig.~\ref{fig:6}(b), 
provide the ground-state diagram in ($r_s$,$N^{-1}$) space in Fig.~\ref{fig:3}. 

%%%%%%%%%< Paragraph 31: Embedding energy: calculation scheme in LSDA >%%%%%%%%%
\subsection{Embedding energy}
\label{Subsec:3D}

In calculating $\delta E$ by use of Eq.~(\ref{eq:11}) at some finite $N$, we 
need $E_0(N\!-\!1,0)$ in addition to $E_0(N,1)$, but for an even number of $N$, 
the ground state of the ($N\!-\!1$)-electron system is necessarily a 
spin-polarized one. Nevertheless, for smoothly connecting to the 
spin-unpolarized ground state at $N\!\to \!\infty$, it is better to calculate 
in the spin-unpolarized situation. In LSDA, by a spin-symmetrized sum of KS 
orbitals by fractionally occupying the orbitals at the Fermi level, we can 
obtain this needed spin-unpolarized ground-state energy $E_0(N\!-\!1,0)$ with 
which we can calculate $\delta E$ for finite-$N$ systems. 

%%%%%%%%%%%%%%%%%%%%%%%%%%%%%%%%%%< Figure 8 >%%%%%%%%%%%%%%%%%%%%%%%%%%%%%%%%%%
\begin{figure}[htbp]
\begin{center}
\includegraphics[scale=0.49,keepaspectratio]{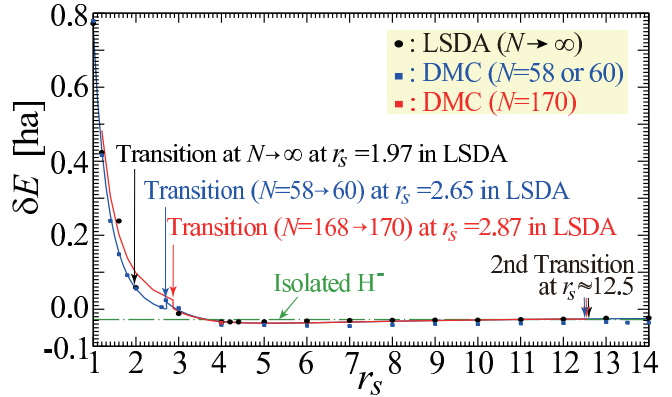}
\end{center}
\caption[Fig.8]{(Color online) Embedding energy $\delta E$ calculated in the 
same situations as those for $n(0)$ in Fig.~\ref{fig:7}. Errors in DMC are 
within the size of square symbols. 
} 
\label{fig:8}
\end{figure}
%------------------------------------------------------------------------------%

%%%%%%%%%< Paragraph 32: Embedding energy: calculation scheme in DMC >%%%%%%%%%%
In DMC, however, we cannot adopt this procedure. Thus we first calculate the 
spin-unpolarized ground-state energy $E_0(N\!-\!2,0)$ in DMC, starting 
with the Slater determinant $\Phi_0({\bm r}_1,\cdots,{\bm r}_{N-2})$ without 
the outermost $s$ orbital. Then we estimate $E_0(N\!-\!1,0)$ by 
\begin{align}
E_0(N\!-\!1,0)\!=\!\frac{N\!-\!1}{N\!-\!2}E_0(N\!-\!2,0)
\left[ 1\!+\!\frac{\alpha_{N\!-\!1}}{(N\!-\!1)^{1/3}}\right ],
\label{eq:15}
\end{align}
with the coefficient $\alpha_{N\!-\!1}$ determined through $E_0(N\!-\!1,0)$ 
and $E_0(N\!-\!2,0)$ in LSDA. We have deduced this approximation scheme by 
considering that the leading term in $E_0(N\!-\!1,0)$ is, in general, 
in proportion to ($N\!-\!1$) due to extensiveness of the total energy as well 
as the subleading term in proportion to $(N\!-\!1)^{2/3}$ due to the 
surface-energy contribution. In CR in which the outermost $s$ orbital is empty 
even for the system with $Z\!=\!1$ and $N$ electrons, instead of 
$E_0(N\!-\!2,0)$, we calculate $E_0(N,0)$ in DMC, with which we estimate 
$E_0(N\!-\!1,0)$ by a similar strategy. 

%%%%%%%%%%%%%%%%%%%%%%%%< Paragraph 33: Embedding energy >%%%%%%%%%%%%%%%%%%%%%%
In Fig.~\ref{fig:8}, $\delta E$ is given as a function of $r_s$, exhibiting 
the $N$ dependence similar to that in $n(0)$ in Fig.~\ref{fig:7}, including 
jumps at CR-SSR transitions. Our results for $\delta E$ agree reasonably well 
with previous ones~\cite{Ref1,Ref2,Ref3,Ref4,Ref5,Ref6,Ref7,Ref8}, though due 
attention to the size dependence was not paid previously. These results of 
$\delta E$ demonstrate that hydrogen is most stably embedded in EG in the form 
of SSR. Thus the concept of SSR is deemed to play a key role in hydrogen 
storage in metals and hydrogen is expected to reside at a site with $r_s\! 
\approx \! 4$ in an inhomogeneous metal. 

%%%%%%%%%%%%%%%%%< Paragraph 34: Anomalous Friedel oscillation.1 >%%%%%%%%%%%%%%
\subsection{Anomalous Friedel oscillation}
\label{Subsec:3E}

Basically, the Friedel oscillation due to the presence of a proton at the 
origin is a concept defined in the bulk system, but if we try to discuss it 
with use of the data in DMC, we need to eliminate the sphere-boundary effect 
from $n(\bm{r})$ obtained in finite-$N$ systems. For this purpose, we adopt the 
following procedure~\cite{Ref8}; we first calculate the charge distribution 
$n_{N,Z}(\bm{r})$ in the finite-$N$ system corresponding to $E_0(N,Z)$ and 
then we estimate $n(\bm{r})$ through the cancellation of the sphere-boundary 
effect by subtracting $n_{N-1,0}(\bm{r})$ from it as
\begin{align}
n(\bm{r}) \approx n_{N,1}(\bm{r})-n_{N-1,0}(\bm{r})+n_0.
\label{eq:17}
\end{align}
In LSDA, we can employ Eq.~(\ref{eq:17}) by using $n_{N-1,0}(\bm{r})$ which 
is obtained in the spin-unpolarized situation simultaneously with $E_0(N\!-
\!1,0)$, but in DMC, only $n_{N-2,0}(\bm{r})$ associated with $E_0(N\!-\!2,0)$ is 
available. Thus we estimate $n_{N-1,0}(\bm{r})$ from this $n_{N-2,0}(\bm{r})$ as
\begin{align}
n_{N-1,0}^{\rm DMC}(\bm{r}) = \lambda^{\beta(r)}\,
n_{N-2,0}^{\rm DMC}(\lambda \bm{r}),
\label{eq:18}
\end{align}
where $\lambda$ is a parameter to correct the difference in the sphere radius, 
defined as $\lambda=[(N\!-\!2)/(N\!-\!1)]^{1/3}$, and the $r$-dependent exponent 
$\beta(r)$ is determined with use of the data for the electron distribution 
in LSDA as 
\begin{align}
\beta(r)=\ln 
\left [n_{N-1,0}^{\rm LSDA}(\bm{r})/
n_{N-2,0}^{\rm LSDA}(\lambda \bm{r}) \right ]
/\ln\lambda.
\label{eq:18a}
\end{align}

%%%%%%%%%%%%%%%%%%%%%%%%%%%%%%%%%%< Figure 9 >%%%%%%%%%%%%%%%%%%%%%%%%%%%%%%%%%%
\begin{figure}[hbtp]
\begin{center}
\includegraphics[scale=0.54,keepaspectratio]{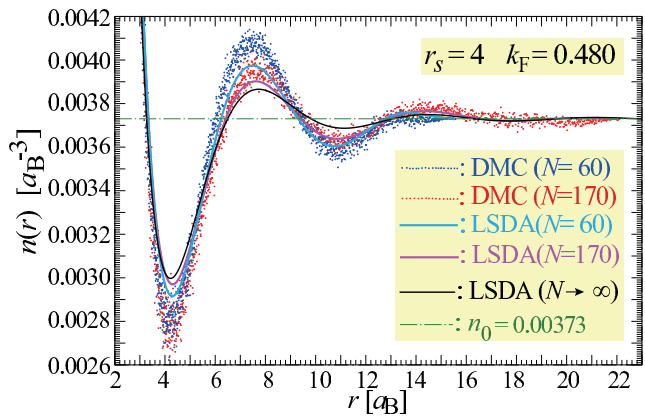}
\end{center}
\caption[Fig.9]{(Color online) Long-range part of $n(\bm{r})$ in both DMC and 
LSDA obtained by the elimination of the sphere-boundary effect.
} 
\label{fig:9}
\end{figure}
%------------------------------------------------------------------------------%

%%%%%%%%%%%%%%%%%< Paragraph 35: Anomalous Friedel oscillation.2 >%%%%%%%%%%%%%%
An example of $n(\bm{r})$ obtained through the above procedure at $r_s\!=\!4$ 
is given in Fig.~\ref{fig:9}, in which we find good convergence for $r\!\lesssim
\!7$ in changing $N$ by the comparison between $N\!=\!60$ and 170, illustrating 
that this procedure doubles the size-convergent range of $r$ as compared with 
that in Fig.~\ref{fig:5}. Admittedly, discrepancy is seen in the oscillation 
amplitude between DMC and LSDA, but overall good agreement and size convergence 
are obtained in the Friedel-oscillation phase, indicating that accurate enough 
information is now available on the node positions in $\delta n({\bm r})$ 
up to $r \!\approx \! 13$. Incidentally the outermost $s$ orbital contributes 
much to the oscillation behavior in Fig.~\ref{fig:9}, assuring its importance 
in the SSR-density region.

\begin{figure}[htbp]
\begin{center}
\includegraphics[scale=0.55,keepaspectratio]{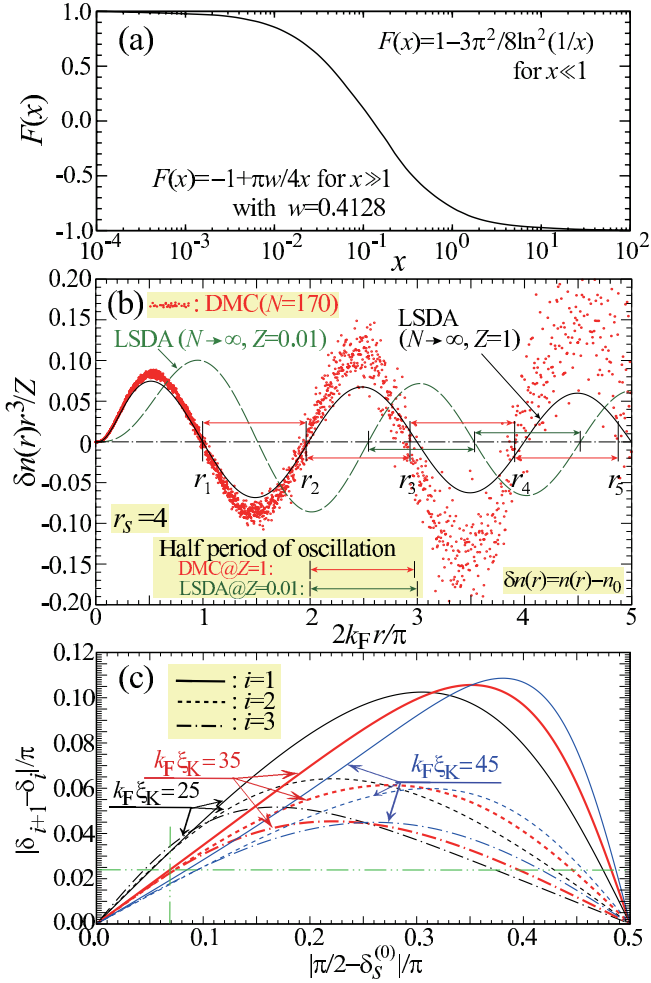}
\end{center}
\caption[Fig.10]{(Color online) (a)Universal scaling function characterizing 
SSR~\cite{Ref12,Ref12b}. (b)Friedel oscillation in SSR in the 
proton-embedded EG at $r_s\!=\!4$, as seen by the plot of $\delta n(\bm{r})
r^3/Z$ in both DMC and LSDA to detect the half period of the oscillation 
in comparison with that in CR obtained in the system with $Z\!=\!0.01$, 
a fictitious tiny charge to ensure that the ground state can be accurately 
obtained in LSDA as a CR state even at $r_s\!=\!4$. (c)First three modulation 
factors in the half period of the anomalous Friedel oscillation, given 
through Eq.~(\ref{eq:16b}) with use of $F(x)$ in (a), plotted as a function 
of the $s$-wave potential scattering phase shift $\delta_s^{(0)}$ 
for various values of $k_{\rm F}\xi_{\rm K}$.
} 
\label{fig:10}
\end{figure}

%%%%%%%%%%%%%%%%%< Paragraph 36: Anomalous Friedel oscillation.3 >%%%%%%%%%%%%%%
In order to unambiguously confirm the emergence of SSR, let us examine 
the Friedel oscillation in $n({\bm r})$ in the light of its 
general behavior, known as~\cite{Ref12a,Ref12,Ref12b,Ref12c}
\begin{align}
%\delta n({\bm r})\!\equiv \!
n({\bm r})\!
\xrightarrow[r \gg k_{\rm F}^{-1}]{}\!n_0\!+\!
\frac{1}{4\pi^2 r^3}\Bigl[&
\cos(2k_{\rm F}r\!-\!3\pi/2\!+\!2\delta_s^{(0)})F(r/\xi_{\rm K})
\nonumber \\
&-\!\cos(2k_{\rm F}r\!-\!3\pi/2)\Bigr ].
\label{eq:16x}
\end{align}
Here only the $s$-wave contribution, which indeed dominates others 
in the present case, is considered and $\delta_s^{(0)}$ is the $s$-wave phase 
shift at the Fermi level produced by the potential scattering without the 
Kondo-resonance effect. In Eq.~(\ref{eq:16x}), $F(r/\xi_{\rm K})\!\equiv\!1$ 
in CR, but in SSR it gradually decreases from $1$ for $r\!\ll\! \xi_{\rm K}$ 
to $-1$ for $r\!\gg\! \xi_{\rm K}$ due to physics of asymptotic 
freedom~\cite{Ref12}. Actually in SSR, $F(x)$ is known to be a universal scaling 
function, as explicitly given in Fig.~\ref{fig:10}(a). 

%%%%%%%%%%%%%%%%%< Paragraph 37: Anomalous Friedel oscillation.4 >%%%%%%%%%%%%%%
By appropriately choosing the branch of $\tan^{-1}x$, we can rewrite 
Eq.~(\ref{eq:16x}) into
\begin{align}
%\delta n({\bm r})\!\equiv \!
n({\bm r})
\xrightarrow[r \gg k_{\rm F}^{-1}]{}n_0-
 \frac{A(r)}{4\pi^2 r^3}
\cos[2k_{\rm F}r\!+\!\delta(r)],
\label{eq:16}
\end{align}
with the amplitude $A(r)$ and the phase $\delta(r)$, given by
\begin{align}
A(r)&=\sqrt{1\!-\!2F(r/\xi_{\rm K})
\cos(2\delta_s^{(0)})\!+\!F(r/\xi_{\rm K})^2},
\label{eq:16a}
\\
\delta(r)&=\tan^{-1}\left [
\frac{1\!-\!F(r/\xi_{\rm K})\cos(2\delta_s^{(0)})}
{F(r/\xi_{\rm K})\sin(2\delta_s^{(0)})}\right ],
\label{eq:16b}
\end{align}
If $\delta_s^{(0)}$ is in the range $(0,\pi/2)$, $\delta(r)$ increases 
gradually from $\delta_s^{(0)}$ for small $r$ to $\delta_s^{(0)}\!+\!\pi/2$ 
for large $r$ in accordance with the change of $F(r/\xi_{\rm K})$. On the other 
hand, if $\delta_s^{(0)}$ is in the range $(\pi/2,\pi)$, $\delta(r)$ 
decreases gradually from $\delta_s^{(0)}$ for small $r$ to $\delta_s^{(0)}\!-
\!\pi/2$ for large $r$. With taking care of such a gradual change in $\delta(r)$, 
we can determine $r_i$ the $i$th zero of $\delta n({\bm r})\,[\equiv\! n({\bm r})
\!-\!n_0]$ by $2k_{\rm F}r_i\!+\!\delta_i\!=\!i\pi\!+\!\pi/2$ with $\delta_i\! 
\equiv \!\delta(r_i)$ and $i\!=\!1,2,3,\cdots$. Then the half period of the 
Friedel oscillation $\Delta_i$ in SSR is given by $\Delta_i\! =\! r_{i+1}\!-
\!r_i\! =\! [1\!-\!(\delta_{i+1}-\delta_i)/\pi]\Delta^{(0)}$ with $\Delta^{(0)} 
(\equiv \!\pi/2k_{\rm F})$ the half period in CR. Because $\delta_{i+1}$ is 
slightly different from $\delta_i$, $\Delta_i$ is modified from $\Delta^{(0)}$ 
by the amount of $(\delta_{i+1}-\delta_i)/\pi$. This anomalous Friedel 
oscillation or {\it the oscillation-period modulation effect} is 
an important consequence of the presence of SSR, but this modulation effect has 
not been well recognized even in the Kondo-physics community due to the 
fact that the effect is totally absent for $\delta_s^{(0)}$ being equal to a 
multiple of $\pi/2$ (see the denominator in Eq.~(\ref{eq:16b})), which happened 
to be assumed in the previous model calculations~\cite{Ref12,Ref12b}. 

%%%%%%%%%%%%%%%%%%%%%%%%%%%%%%%%%%< Figure 10 >%%%%%%%%%%%%%%%%%%%%%%%%%%%%%%%%%
%------------------------------------------------------------------------------%

%%%%%%%%%%%%%%%%%< Paragraph 38: Anomalous Friedel oscillation.5 >%%%%%%%%%%%%%%
In Fig.~\ref{fig:10}(b), we plot $\delta n(\bm{r})r^3/Z$ as a function of 
$2k_{\rm F}r/\pi$ at $r_s\!=\!4$ to check whether the modulation effect exists 
or not in our first-principles calculations. In DMC, the first four zeros of 
$\delta n({\bm r})$ in units of $\Delta^{(0)}$, $2k_{\rm F}r_i/\pi$, are given 
by $0.994\pm0.005$, $1.970\pm0.008$, $2.949\pm0.013$, and $3.927\pm0.019$ with 
the errors estimated by the distribution of data points around 
$\delta n({\bm r}\!)=\!0$ at each $r_i$. Then $(\delta_{i+1}\!-\!\delta_i)/\pi$ 
the modulation factors for $i=1,2$, and $3$ are, respectively, obtained as 
$0.024\pm0.005$, $0.021\pm0.009$, and $0.022\pm0.016$ in which the errors are 
estimated by the inclusion of covariance between $r_i$ and $r_{i+1}$. Those 
results, at least definitely those for $i=1$ and 2, assure the existence of 
the shortening of the Friedel-oscillation period, confirming the emergence of SSR. 

%%%%%%%%%%%%%%%< Paragraph 39: Estimation of the Kondo temperature >%%%%%%%%%%%%
Once the data for $\{r_i\}$ are known, we can independently calculate the 
modulation factors through Eq.~(\ref{eq:16b}) as a function of $\xi_{\rm K}$ and 
$\delta_s^{(0)}$ with using $F(x)$ in Fig.~\ref{fig:10}(a). The results are shown 
in Fig.~\ref{fig:10}(c), from which we find that our DMC data for the modulation 
factors (actually the shortening factors in this case) for $i=1-3$ agree very 
well with those obtained at $\xi_{\rm K}\approx 35/k_{\rm F}=73$ and 
$\delta_s^{(0)}\!=\!0.86\!\times\! (\pi/2)$ (see the horizontal and vertical 
double-dotted-dashed lines in Fig.~\ref{fig:10}(c)), indicating that by 
quantitatively analyzing first few modulation factors, we {\it can determine both 
$\xi_{\rm K}$ and $\delta_s^{(0)}$ uniquely}, even if only the data for the system 
size much shorter than $\xi_{\rm K}$ are available. This determination is made 
possible due to the fact that $F(r/\xi_{\rm K})$ changes most rapidly at $r \approx 
\xi_{\rm K}/10$. By using $\xi_{\rm K}$ thus determined and the Fermi velocity 
$v_{\rm F}$, the Kondo temperature $T_{\rm K}$ is estimated by $T_{\rm K}=
v_{\rm F}/\xi_{\rm K}\approx 0.0066$ Hartree $=2,100$K with about 10\% errors, 
revealing that this is an astonishingly high-$T_{\rm K}$ system. With such a 
high $T_{\rm K}$, the system will not exhibit the prominent Kondo effects such 
as the $\ln T$ anomaly in the experiment for $T$ around the room temperature 
or below; it just behaves as a usual Fermi liquid~\cite{Ref16b}.

%%%%%%%%%%%%%%%%%%%%%%%%%%%%%%%%%< Figure 11 >%%%%%%%%%%%%%%%%%%%%%%%%%%%%%%%%%%
\begin{figure}[htbp]
\begin{center}
\includegraphics[scale=0.57,keepaspectratio]{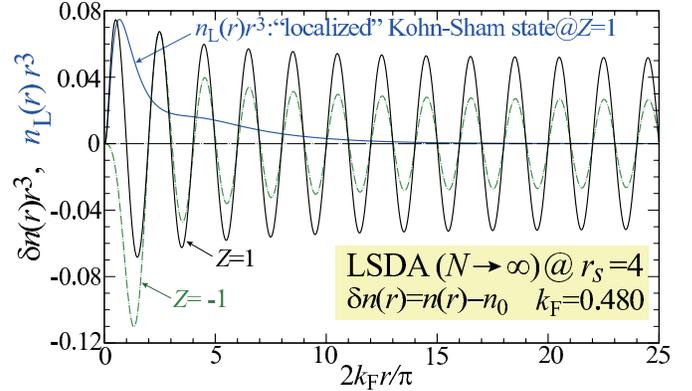}
\end{center}
\caption[Fig.11]{(Color online) Plot of $\delta n(\bm{r})r^3$ in LSDA in the 
bulk system with both $Z=1$ and $-1$ at $r_s=4$. The contribution from the 
``localized'' KS state obtained at $Z=1$, $\delta n_{\rm L}(\bm{r})r^3$, is 
also given and seen as playing a role of an envelope density. 
} 
\label{fig:11}
\end{figure}
%------------------------------------------------------------------------------%

%%%%%%%%%%%%%%< Paragraph 40: Anomalous Friedel oscillation in LDA >%%%%%%%%%%%%
The shortening effect in the Friedel-oscillation period is also found in LSDA, 
as seen in Fig~\ref{fig:11}, in which $\delta n(\bm{r})r^3$ is plotted for both 
$Z=1$ and $-1$ in the bulk system at $r_s=4$. For large enough $r$ outside the 
SSR binding radius $\xi_{\rm K}$, the Friedel-oscillation phase for $Z=1$ 
coincides with that for $Z=-1$; the result with $Z=-1$ is plotted to represent 
the behavior for the fictitious very tightly bound $H^-$ ion confinement state 
in which the anomalous Friedel oscillation is absent. In LSDA, the modulation 
of the Friedel oscillation is brought about by the contribution from the 
``localized'' density $n_{\rm L}(\bm{r})$. Incidentally, the oscillation-period 
shortening factors in LSDA are much smaller than those in DMC; $(\delta_{2}\!-\!
\delta_1)/\pi=0.0133$ and all others are less than 0.004, indicating 
$\delta_s^{(0)}\!\approx \!0.014\!\times\! (\pi/2)$ in LSDA. (We note that 
$\delta_s(k_{\rm F})$ in Fig.~\ref{fig:4}(b) is equal to $\delta_s^{(0)}\!+\!
\pi/2\approx 0.507\pi$.) We can easily understand the reason for this large 
difference in $\delta_s^{(0)}$ between DMC and LSDA; because this phase shift 
is directly connected with the wave function, its accurate value will not be 
obtained by LSDA in which the wave function in the fictitious non-interacting 
system is qualitatively  different from the true correlated SSR wave function. 
Due to $\delta_s^{(0)}\approx  0$ in LSDA, we cannot employ the diagram in 
Fig.~\ref{fig:10}(c) to determine $\xi_{\rm K}$ very accurately. Therefore 
we estimate $\xi_{\rm K}$ from the extent of $n_{\rm L}(\bm{r})$, which is 82, 
giving $T_{\rm K}$ to be about 1,900K. In relation to $n_{\rm L}(\bm{r})$, 
$\varepsilon_{\rm BS}$ the binding energy for this ``localized'' KS state in 
LSDA is given as 0.0115 Hartree = 3,600K, which is about twice as large as 
$T_{\rm K}$. Thus, although $\varepsilon_{\rm BS}$ has no direct physical 
meaning, this quantity seems to be a good measure for the magnitude of 
$T_{\rm K}$. In our calculations in LSDA, the values for $\varepsilon_{\rm BS}$ 
are 97 K, 2,200K, 3,500K, 2,200K, and 720K for $r_s=2.07$, 3, 6, 8, and 10, 
respectively, indicating that in the majority of the SSR region, namely, for 
$3 \lesssim r_s \lesssim 8$, we may expect $T_{\rm K}$ to be well beyond 1,000K.

%%%%%%%%%%%%%%%%%%%%%%%%%%%%%%%%%[ Section 4 ]%%%%%%%%%%%%%%%%%%%%%%%%%%%%%%%%%%
%%%%%%%%%%%%%%%%%%%%%< Paragraph 41: Discussion and Comments >%%%%%%%%%%%%%%%%%%
\section{Discussion}

Five comments on the present work are in order: 

(i) In SSR, we find a good semi-quantitative 
agreement between LSDA and DMC, but this is by no means fortuitous, because 
this can be understood by the long-range nature of $\xi_{\rm K}$ which makes the 
density variation associated with the SSR state slow, validating the use of LSDA 
for the calculation of $n({\bm r})$. 

(ii) From our present perspective, we may regard our previous study in LSDA on 
the spin-polarized ground states for second-period atoms in Periodic 
Table~\cite{Ref20} as a successful extension to multi-channel Kondo 
systems~\cite{Ref10}, in which the Hund's-rule coupling plays a crucial role 
in producing the spin-polarized ground states. 

(iii) In a short term, the immediate next target of research is a hydrogen 
molecule H$_2$ immersed in EG~\cite{Ref3} to pursue a new concept in chemical 
bonding~\cite{Ref24,Ref25} in metals in the light of SSR. In fact, Bonev and 
Ashcroft~\cite{Ref23} have already found an interesting bistability 
between paired and unpaired states for $r_s>3.2$. In a longer term, we can 
expect fruitful research on new aspects in the Ruderman-Kittel-Kasuya-Yosida 
(RKKY) interaction~\cite{Ref10} and dilute magnetic semiconductors~\cite{Ref26}. 

(iv) If the number of immersed protons is increased up to a macroscopic level 
and those protons are arranged into a lattice in a metal with the SSR-density 
region, we may obtain a Kondo lattice with high $T_{\rm K}$. Furthermore, if the 
lattice constant is so arranged as to tune the RKKY interaction to be about 
the same as $T_{\rm K}$ for realizing the quantum-critical situation in the 
Doniach phase diagram~\cite{Ref40}, then as indicated in Fig.~\ref{fig:1}, 
we may obtain high-$T_c$ superconductivity in the spin-fluctuation 
mechanism in a metallic hydrogen alloy. In this regard, the 
transition-metal-hydride system such as TiH$_2$ and 
ZrH$_2$~\cite{Ref41,Ref42,Ref43}, usually used for batteries, might be a 
promising candidate, although it seems that the metallic electron densities 
in the metal hydrides so far synthesized are too high for our purpose.

(v) The basic reason for about hundred-times increase of $T_{\rm K}$ and 
consequently $T_c$ in the hydrogen system compared with those in the 
$f$-electron heavy-fermion systems is the overall increase of energy scales 
as seen by the large difference in the magnitudes of 1$s$ and 4$f$ energy levels. 
For the same reason of increased energy scales, superconductivity with 
$T_c$ over 100K has been discussed in solid hydrogen at pressures of about 
500GPa in the conventional phonon mechanism~\cite{Ref31,Ref32,Ref32a,Ref32b}. 
Very recently, stimulated by the experimental result~\cite{Ref33}, similar 
discussions are made on H$_2$S~\cite{Ref34,Ref36} as well as H$_3$S~\cite{Ref35} 
at pressures of about 200GPa. Note that the metallic densities in those 
systems are found to be in the CR region, namely, $r_s<2$ (or typically $r_s 
\approx 1.4$). Therefore, our present proposal has nothing to do with those 
conventional theories for superconductivity. We also emphasize that high 
metallic densities realized at such very high pressures are not needed in our 
proposal, indicating that high-$T_c$ superconductivity is expected to occur 
even {\it at ambient pressure}, contrary to the case of solid H$_2$ or 
sulfur hydrides. 

%%%%%%%%%%%%%%%%%%%%%%%%%%%%%%%%%[ Section 5 ]%%%%%%%%%%%%%%%%%%%%%%%%%%%%%%%%%%
%%%%%%%%%%%%%%%%%%%%%%%%%%%< Paragraph 42: Summary >%%%%%%%%%%%%%%%%%%%%%%%%%%%%
\section{Summary}

In conclusion, by employing both DMC and LSDA, we have revealed the emergence 
of SSR in the proton-embedded electron gas by the confirmation of an anomalous 
Friedel oscillation characteristic to the Kondo-like spin-singlet state with 
quantitatively determining $T_{\rm K}$ and emphasized its stability in 
embedding hydrogen into the electron gas. Our work necessitates to modify the 
paradigm of CR in metallic screening to a point charge of $Z$ in the sense 
that, depending on the metallic electron density and $Z$, SSR takes the place 
of CR. This work also provides a first concrete example to show how the KS 
orbitals behave to represent the exact $n({\bm r})$ in strongly-correlated 
electron systems such as those describing Kondo physics, even though they are 
not always physically relevant. Finally, this work discloses a long-sought 
high-$T_{\rm K}$ system, opening an unexpected route to room-temperature 
superconductivity in a Kondo lattice made of protons and motivating people 
engaged in hydrogen-based physics, chemistry, and technology to synthesize 
hydrogen alloys exhibiting high-$T_c$ superconductivity at ambient pressure.

%%%%%%%%%%< Acknowledgement >%%%%%%%%%%
\acknowledgments

Y.T. thanks A. Savin and E. K. U. Gross for valuable discussions. This work is 
partially supported by Innovative Area "Materials Design through Computics: 
Complex Correlation and Non-Equilibrium Dynamics" (No. 22104011) from MEXT, Japan. 

%%%%%%%%%%<  References  >%%%%%%%%%%

%%%%%%%%%%%%%%%%%%%%%%%%%%%%%%%%%%%%%%%%%%%%%%%%%%%%%%%%%%%%%%%%%%%%%%%%%%%%%%%%
\end{document}